\documentclass[article]{revtex4}
\usepackage{graphicx}

\def\ord{{\cal O}}

\def\depth{depth}
\def\Depth{Depth}
\def\D{{\cal D}}
\newcommand{\PP}{{\bf P}}
\newcommand{\NC}{{\bf NC}}
\newcommand{\NP}{{\bf NP}}
\newcommand{\PS}{{\bf PSPACE}}
\newcommand{\AC}{{\bf AC}$^0$}
\newcommand{\CC}{{\bf CC}}

\begin{document}

\title{Complexity, parallel computation and statistical physics}
\author{J. Machta}
\email[]{machta@physics.umass.edu}
\address{Department of Physics,
University of Massachusetts,
Amherst, MA 01003-3720}
\begin{abstract}

The intuition that a long history is  required for the emergence of complexity in natural systems is formalized using the notion of {\em \depth }.  The  \depth\ of a system is defined in terms of the number of parallel computational steps needed to simulate it. 
\Depth\ provides an objective, irreducible measure of history applicable to systems of the kind studied in statistical physics. It is argued that physical complexity cannot occur in the absence of substantial \depth\ and that \depth\ is a useful proxy for physical complexity.  The ideas are illustrated for a variety of systems in statistical physics.  

\end{abstract} 

\maketitle

\section{Introduction}

The history of the Universe is marked by increasing complexity manifest on many scales:
cosmological, geological, biological and social.  Since ancient times, humankind has sought to
explain why complexity increases. Theological explanations have given way to specific
mechanisms--Darwinian evolution in biology and  gravitational amplification of primordial
fluctuations in cosmology-- but there is still very little general understanding of why or how
complexity increases.  Part of the difficulty lies in the fact that there is no agreed upon definition
of  ``complexity."  How can we explain why something increases if we don't know what it is?  We can for the most part agree on a hierarchy of complexity that places homogeneous equilibrium systems at the bottom of the ladder and biological systems near the top.  While many definitions of complexity have been offered \cite{Grass86,Benn87, Benn88,LlPa88,Gell95,CrSh99} none has gained wide acceptance and it may not be possible to give a definition of complexity that captures all of its manifestations.  Justice Potter Stewart's famous definition of pornography may be equally applicable to complexity, ``I shall not today attempt further to define the kinds of  material I understand to be embraced with that shorthand description . . . But I know it when I see it."  Although  the complexity of natural systems is a chief concern of this paper, we will be content with Potter's definition and instead seek to define a proxy quantity whose presence is required for the  emergence of complexity. 

An obvious fact about complexity is the basis for this paper.   {\em The emergence of complexity requires a long history.}  A long history is a central feature of life on Earth and is manifested, for example, in the depth of the phylogenetic tree.  Although the passage of time is necessary for the emergence of complexity, time alone is not sufficient.  For example,  an isolated container filled with a gas will remain in equilibrium indefinitely with no increase in complexity.  The word ``history'' implies more than the passage of time and, in this sense, an equilibrium system has very little history. The purpose of this paper is to introduce a formal, irreducible measure of history and to show that it captures some of the intuitive properties associated with complexity.  The new quantity  is defined using concepts borrowed from theoretical computer science and is applied to systems in statistical physics.  It is irreducible in the sense that a much stricter measure would yield the uninteresting result that no physical process generates a long history.

Charles Bennett has explored the idea that complexity emerges only
after a long history~\cite{Benn87, Benn88,Benn90,Benn95}.  He proposed {\em logical depth} as an appropriate way to measure history and physical complexity.  The logical depth of a system state is the execution time on a Turing machine needed to produce a description of that state starting from a short program.  Logical depth is, roughly speaking, the total number of elementary computational operations needed to produce a description of the state of the system.   This paper was inspired by Bennett's ideas and starts from the same two premises: (1) a long history is a prerequisite for the emergence of complexity and (2) computation is the correct domain for measuring the length of a history.  Other studies motivated by Bennett's ideas can be found in Refs.\ \cite{JuLaLu94,May04}.

There are two key differences  between Bennett's logical depth and the depth measure presented here.  The first is the choice of computational resource used to measure the length of a history.  Bennett chooses the total number of Turing operations~\footnote{Although logical depth usually refers to time on a Turing machine, in \cite{Benn87}, Bennett suggests that time on a three dimensional cellular automata might be an appropriate way to define depth.} whereas I propose to measure the length of a history as the number of {\em parallel} computational steps required to simulate the system.  I will call this measure physical \depth\ or parallel \depth\ when it is necessary to distinguish it from other meanings of the word depth or simply \depth\ when no ambiguity results.  Because many logical operations done simultaneously are counted as a single parallel step, it is possible that the parallel \depth\ of a large system is small even if the logical depth is large.  The second difference between the two definitions is that logical depth refers to individual system states whereas parallel \depth\ applies to ensembles or probability distributions over system states.  Parallel \depth\ is thus the running time of a {\em Monte Carlo} algorithm that generates a typical state of the system.  The probabilistic character of physical \depth\ is similar to other quantities, such as entropy or temperature, defined in statistical physics.

A difficulty in developing a general theory of complex systems is that complexity is an epiphenomenon.  Cosmological complexity is manifest in the organization of visible matter into stars, galaxies and clusters of galaxies but these visible aspects of the Universe account for a small fraction of the stuff of the Universe, most of which is dark matter and dark energy, presumably devoid of complexity.  Biological complexity resides in a thin film covering the Earth and the biosphere survives by capturing a miniscule fraction of the output of the Sun. The farther up the ladder of complexity one looks, the more tenuous, fragile and contingent are the phenomena.  The fact that we have found no evidence for life except on Earth tends to confirm the view that complexity is a rare and accidental feature of the Universe and not a pervasive or necessary  consequence of physical laws.  It is difficult to imagine that a robust, simple measure could be sensitive to the existence of complexity and could, for example, distinguish a lifeless planet from one that contains life.  Nonetheless,  physical \depth\ may be large even for systems where complexity is an epiphenomenon.   For example, it may be that the \depth\ of the solar system is dominated by the biosphere.

The grand questions surrounding complexity are a primary motivation for the work reported here.  However, when stripped of these motivations what remains is a study of the most efficient strategies for simulating natural systems on massively parallel computers.  Some of these strategies may be directly or indirectly useful in computational science.

The organization of the paper is as follows. The next section informally motivates the notion of  \depth\ and sets the stage for the more formal definition given in Sec.\ \ref{sec:def}.  \Depth\ is defined in the language of parallel computing and computational complexity theory, briefly introduced in Sec.\ \ref{sec:cc}, and applies to the domain of statistical physics, briefly introduced in Sec.\ \ref{sec:sp}.  Section \ref{sec:examples} considers the \depth\ of several well-studied systems in statistical physics.  The paper closes with a discussion in Sec.\ \ref{sec:disc}.

\section{Measuring history}
\label{sec:intuitive}

The central hypothesis of this paper is that the emergence of complexity requires a long history. Time, as  understood in physics and everyday life, is surely required for the emergence of complexity. However, the passage of time only rarely leads to increasing complexity.  \Depth\ is a logical measure of history that is stricter than physical time. \Depth\ is the minimum number of computational steps needed to simulate a system state.   It is irreducible in the sense that a stricter definition would make it impossible for any (classical) system to have much \depth. 

The shortcomings of
physical time as a certification of a long history and the suitability of a measure based on parallel computation are best seen by considering some examples.  First consider a system of the type studied in thermodynamics, an isolated sample of a gas in a fixed volume.  Suppose that the system is large enough to be composed of a very large number of molecules but small enough that gravity does not play a role.  Such a system approaches equilibrium and, once in equilibrium, its behavior is dull and monotonous; time passes but the statistical properties of the system do not change and nothing of interest happens.  \Depth\ should reflect this observation and the \depth\ of an equilibrium system should not depend on how long it has remained in equilibrium but only on how many steps are needed to reach equilibrium by some efficient process. Equilibrium states, particularly equilibrium states at very high or very low temperatures, are near the bottom of the hierarchy of complexity.   

\begin{figure}
\includegraphics[width=5in]{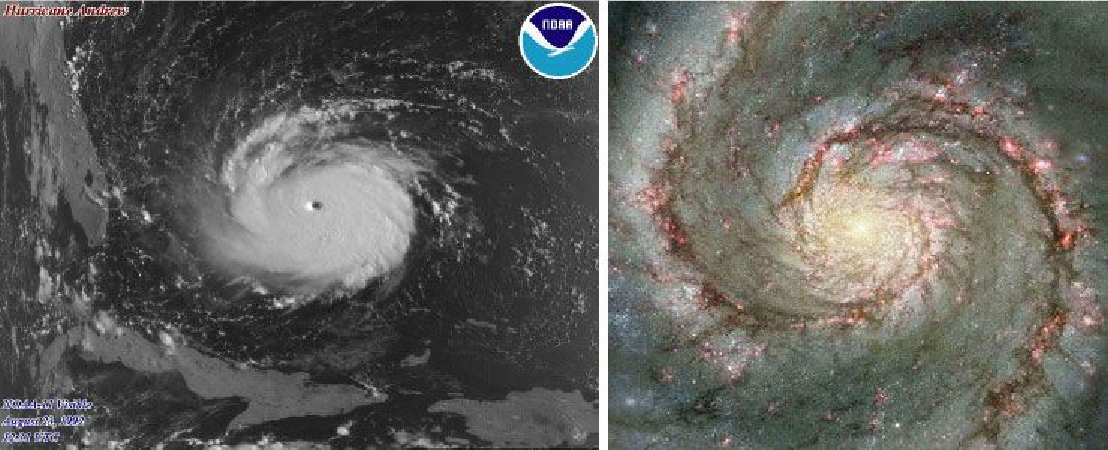}
\caption{Hurricane Andrew approaching Florida (left) and the spiral galaxy M51 (right). Courtesy of NOAA and NASA, respectively.}
\label{fig:hurricane_galaxy}
\end{figure}

Next, let's compare two somewhat more complex systems, a hurricane and a spiral galaxy, shown in Fig.\ \ref{fig:hurricane_galaxy}. Both are rotating structures with a superficially similar appearance but otherwise the physics
involved is quite different.  The hurricane's rotation is the product of the release of latent heat, the consequent lifting of warm air and the concentration of the general rotation of the atmosphere around a deep pressure minimum. 
The time for hurricane formation is measured in days.  The concentration of rotation in
a spiral galaxy is due to the gravitational collapse of the
primordial gas from which the galaxy forms.  The time for galaxy formation is hundreds
of millions of years.  Intuitively, both galaxies and hurricanes represent roughly
comparable, intermediate levels of complexity; they are self-organized structures displaying
much more complexity than an equilibrium gas but much less complexity than
biological systems.  Both hurricanes and galaxies are at the current limits of our abilities to do
reasonably realistic computer simulations.  If these two systems are comparable in
complexity why does it take so much longer to make a galaxy?  It is not because so much
more is happening but rather because galaxy formation occurs on a vastly greater scale than
hurricane formation, a scale measured in millions of light years rather than hundreds of
kilometers.  Thus the time scale for galaxy formation is bounded by a 
communication time at the speed of light of millions of years.   Atmospheric disturbances
propagate more slowly in a hurricane but the distances are much less so the
deep pressure minimum of a hurricane can be set up in a matter of days.  We would not want to say
that the galaxy has a history that is ten billion times longer than a hurricane implying the
potential for vastly more complexity.   The observation that these two systems have comparable complexity leads to two related conclusions. First, the clock measuring \depth\ should be ticking more slowly for the galaxy than the hurricane and second, communication time should be discounted in the measurement of \depth.  Though time is required for the emergence of complexity, it does not follow that a system that changes more slowly is more complex or that complexity increases just because signals are propagating over large distances.

A coarse-grained view is implicit in the statement that a galaxy and a hurricane are of comparable complexity.  In this context a galaxy  is really just a gravitating mass distribution  and its description does not extend down to the level of individual stars, planets, atmospheres or biospheres.  If  these subsystems are incorporated in the model in a detailed way a galaxy would be far more complex than a hurricane and might include hurricanes and even intelligent life \footnote{Mathematically tractable examples of a reduction in computational complexity accompanying coarse-graining are seen in one-dimensional cellular automata~\cite{IsGo94}.}.  The \depth\  of a natural system cannot be defined without first specifying a reasonably independent set of degrees of freedom.  That being said,  the definition of  \depth\ should not to change much when more fundamental degrees of freedom that do not have much complexity are included in the description of the system.  For example, the hurricane should not have a substantially greater \depth\  if it is described at a molecular level rather than a hydrodynamic level even though the number of degrees of freedom is much greater for the molecular description. Unlike the galaxy where a fine-grained view may radically increase complexity and \depth, the fine-grained view of the hurricane reveals gases that are locally near equilibrium and contribute little to the complexity or \depth\ of the whole.  The \depth\ of such a system can be nearly independent of the level of coarse-graining if  \depth\ is defined in terms of parallel computation with a number of processors that scales up appropriately with the number of degrees of freedom of the system.

Given a faithful, quantitative description of a system, \depth\ is defined as the number of steps required to generate a description of a typical system state starting from a simple beginning.  A central hypothesis of this paper is that these steps should be measured on a parallel digital computer using the most efficient possible algorithm.  The tick of the clock measuring the history of a system is a logical step in this {\em simulation} of the system, rather than an observable of the system itself. The simulation that generates the system state is, at least in principle, physical since computers are physical devices but  the path taken by the computer may not be closely related to the natural dynamics of the system.  The best parallel algorithm for simulating an equilibrium gas, hurricane or galaxy will take advantage of shortcuts unrelated to the way these systems actually evolve. 

Measuring \depth\ requires both natural science and computational science.  Experimental measurements and observations are required to verify that a model of a natural system is correct and sufficiently accurate.  Computational science is needed to develop the best ways to simulate the model.   Since experimental measurements are the ultimate arbiter of computational models, \depth\ is, in some sense, a physical property though the connection to experiment may be indirect. However, if \depth\ deserves the status of a physical property, we must show that it is unique, well-defined and not tied to a particular computational model.

If \depth\ is to be a unique measure, some particular algorithm for simulating a system must be
chosen.  We should not credit a system with having a long history and a high potential for
complexity just because we have chosen an inefficient simulation method that
requires many steps.  Thus, \depth\ should be defined with respect to the most
efficient method for simulating the system, that is, the method that requires the fewest
steps.  The algorithms used in the simulation need not correspond to the physical dynamics of the system so long as the end product is a faithful representation of the state of the system.  In Sec.\ \ref{sec:sp} we will discuss examples of accelerated parallel algorithms.   If parallelism allows typical states to be generated in significantly fewer steps than are required by the system's physical dynamics, the system has less history than is naively apparent. 

In principle, it is not generally possible to know the most efficient means for solving computational problems so the requirement of using the
most efficient algorithm makes \depth\ uncomputable.  In
practice, given enough experience simulating a class of systems we can have
some assurance that revolutionary improvements are unlikely so that conclusions drawn from the current state of the art are likely to be of lasting value.  The lack of finality in the measurement of \depth\ is no more disturbing than the lack of finality in any scientific theory.  Indeed, one might argue that as long as the best algorithm for simulating a system is not known, the system is not fully understood.  In any case, existing algorithms set upper bounds on \depth.

In the present work we consider discrete, classical computation, which is also the standard model in computational complexity theory.  The broadest possible view of simulation would consider all physical ways of arriving at a good representation of a system state including classical analog computation and quantum computation.  Discrete, classical computation provides a way to quantify history as a number of computational steps and is well-understood.  However, analog computation~\cite{BlCuShSm,Si97} or quantum computation~\cite{ChNi00} might ultimately prove to be a more  appropriate foundation for understanding physical complexity.  Lurking behind the choice of classical, digital computation is the physical Church-Turing hypothesis~\cite{Wolf85} that can be paraphrased as, ``Any physical process can be efficiently simulated on a digital computer.''

\begin{figure}
\includegraphics[width=5in]{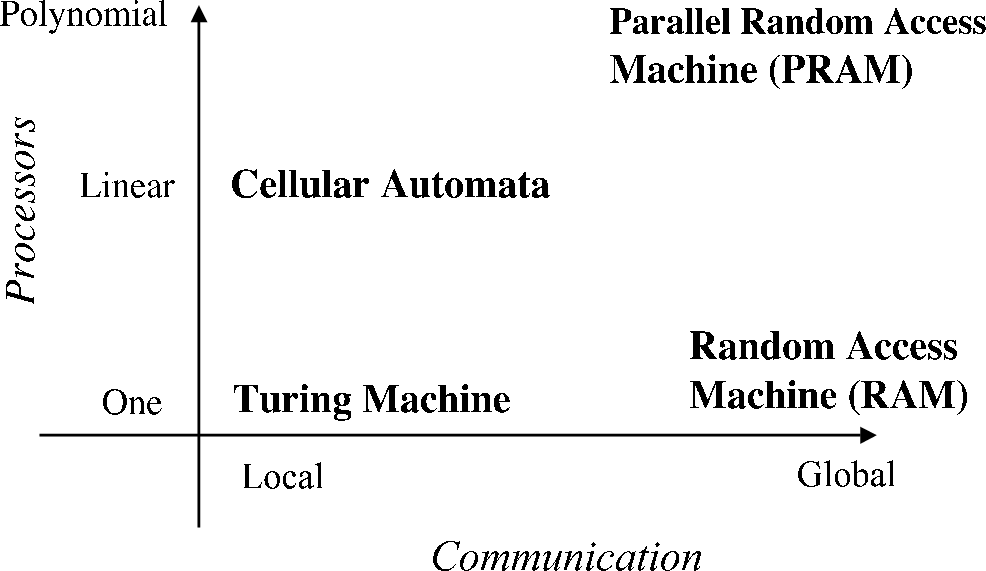}
\caption{Four fundamental models of computation distributed in the horizontal direction according to whether they have local or global communication and in the vertical direction according to the degree of parallelism.}
\label{fig:models}
\end{figure}

Having narrowed the discourse to efficient simulations using digital computers it is still necessary to decide on the appropriate computational resource to associate with history and \depth.  Different models of digital computation naturally lead to different definitions of a computational step. Thus, the discussion can be cast in terms of choosing a model of computation for which the natural notion of an elementary step is best suited to the purpose of measuring history in the natural world.  The two main choices to be made are between sequential and parallel computing and between local and non-local communication.  For parallel computing there is an additional question of how many processors to allow.  Possible choices are reflected in the following fundamental models of computation-- the Turing machine, cellular automata (CA), the random access machine (RAM) and the parallel random access machine (PRAM).  These devices are shown in Fig.\  \ref{fig:models} on a two-dimensional diagram that classifies them according to number of processors and constraints on communication.  
 
The original model for fundamental investigations in the theory of computing was the Turing machine.  It has one processor with a finite number of internal states that moves along a one-dimensional data tape of arbitrary length.  An elementary step in a Turing  computation consists of changing the state of the head,  reading and writing to the tape and then moving one unit to the left or right along the tape.  Turing machines allow neither parallelism nor long range communication.  Turing machines are physically realizable and computationally universal. However, their lack of parallelism and non-local communication means that simulating most physical processes on a Turing machine will require a number of steps that increases more rapidly than the physical time of the process that is simulated.  

Cellular automata have many simple processing elements arranged on a lattice with  communication between nearest neighbor processors.  In a single step, each processor reads the state of its nearest neighbors, carries out a simple logical operation based on that information and makes a transition to a new internal state.  Cellular automata are effective simulators of many physical processes~\cite{Wolf94,Wolf02} and they
are attractive candidates for measuring history because their parallelism and locality most closely mirror the physical world.   Time in a CA simulation is often proportional to physical time.  Cellular automata time combines communication time and processing time in a way similar way to the physical world. 

The RAM and PRAM differ from the Turing machine and cellular automata by allowing non-local communication.  The RAM is an idealized and simplified version of the  ubiquitous desktop computer.  It consists of a single processor with a simple instruction set that communicates with a global random access memory.  In an elementary step, the processor may read from one memory cell, carry out a simple computation based on the information in the cell and its own state and then write to one memory cell. The definition of ``time'' on a RAM presumes that any memory cell can be accessed in unit time.  Thus, the physical time required for a single step of a RAM must ultimately increase at least as the cube root of the number of memory elements due to the finiteness of memory density and signal propagation speed.   The RAM is the customary way of thinking about {\em computational work} or the number of elementary operations needed to carry out a computation.    Since the RAM is reasonable approximation to a single processor workstation, the conventional way to compare the efficiency of algorithms is in terms of time on a RAM.   

\begin{figure}
\includegraphics[width=5in]{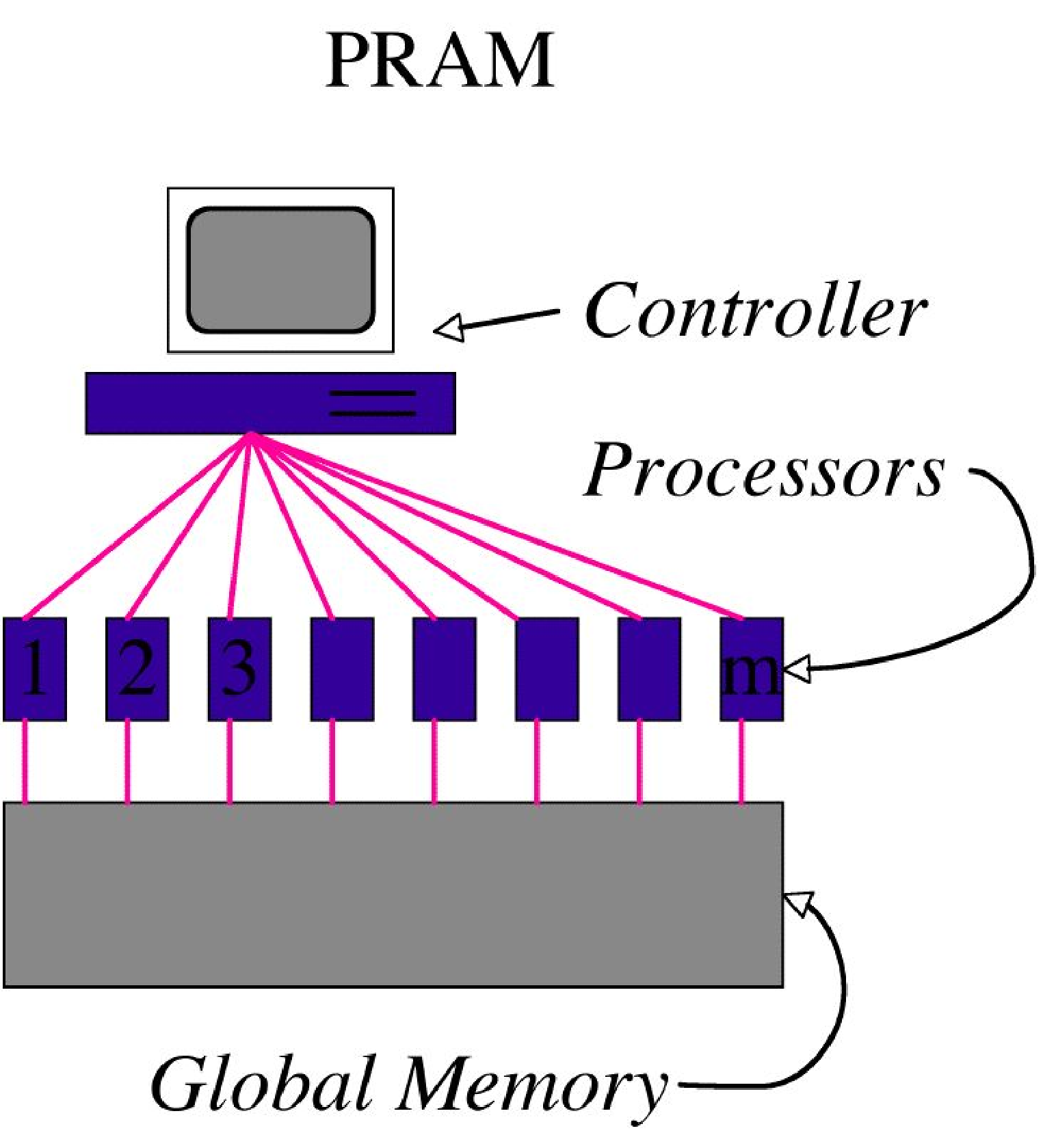}
\caption{The parallel random access machine (PRAM) with $m$ processors, a global random access memory and  an input/output/controller device.}
\label{fig:pram}
\end{figure}

The PRAM is an idealized model of parallel computation with global communication.   The PRAM consists of many identical processors all connected to a single global random access memory and an input/output/controller as shown in Fig.\ \ref{fig:pram}.  The processors are the same as the  processor of a RAM, each is a stripped down microprocessor.  The number of processors is conventionally allowed to grow polynomially (as a power) of the problem size.    As in the case of the RAM, in a single step each processor may read from one memory cell, carry out a simple computation based on the information in the cell and its own state, and then write to one memory cell.  The shared global  memory effectively allows any two processors to communicate with one another in a couple of steps. Additional rules are needed to determine what happens when two processors attempt to write to the same cell.

All of the devices described above are computationally universal meaning that each can simulate the other in a number of steps that differs only polynomially.  The Turing machine and the PRAM are at opposite extremes among computationally universal, discrete classical devices.  The Turing machine does the least in an elementary step and the PRAM the most.   Bennett's original suggestion was to use a universal Turing machine simulation to measure history and to count the number of elementary Turing operations.  I  propose using the PRAM instead.

What are some of the reasons to focus on parallel time (either PRAM or CA) instead of sequential time (either RAM or Turing)?  Consider again a molecular  gas with short range interactions.  The sequential time for simulating such a system for physical time $t$ scales as the product of the size of the system and the elapsed physical time.  However, by using a number of processors proportional to the number of molecules, $N$ the parallel time is independent of $N$.  The parallel time needed to generate an equilibrium state of a gas is nearly independent of $N$ (except at critical points) whereas the sequential time is at least proportional to $N$.  Parallel time is here a better measure of the length of the history of the system and more clearly reflects the potential for complexity--larger samples are not more complex just because they require more computational work to simulate. Of course, as is commonly done in measuring the dynamical properties of Monte Carlo algorithms in
statistical physics, we could measure the sequential time and then simply divide by the number
of molecules in the system to obtain an intensive measure that is independent of system size. 

For homogeneous systems, it might not matter whether \depth\ is
defined in terms of parallel time or sequential time divided by the number of degrees of freedom
but for systems composed of qualitatively different parts, these choices can lead to very different results. For example, consider a toy solar system, consisting of a Sun and an Earth with a biosphere.  A typical state of the Sun has very
little history since its hot gases provide no repository for long term 
memory except of a trivial kind resulting from global conservation 
laws.  A fined-grained parallel simulation of a typical state of the Sun would require a huge number of
processors but relatively few parallel steps.  To carry out such a simulation we would have to
understand the dynamical processes in the Sun with its current composition and then let
the model Sun run long enough to reach a steady state uncorrelated with the initial state of the
simulation.  Most of the physical and computational knowledge needed for actually carrying out
the simulation is already in hand.   On the other hand, obtaining a statistically valid picture
of a several billion year old biosphere would presumably require a simulation covering billions of years since many of the arbitrary choices occurring at the beginning of life have been preserved to the present era.  Although we are very far from knowing how to simulate a biosphere efficiently, it is
plausible to conclude that the \depth\ of the biosphere is far greater than the
\depth\ of the Sun. On the other hand, if the biosphere and the Sun are to be simulated to the
same resolution, the computational work of simulating the Sun is almost certainly larger just due to its much greater size. A  crude comparison of 
computational resources needed to simulate each system that ignores nearly all of the real issues
is obtained from the product of the mass of each subsystem and a time scale to reach a
statistically valid snapshot of its present state. The mass of the Sun is $10^{30}$kg and we can
liberally estimate a time scale or ``\depth'' of a million years for reaching a steady state,
multiplying the two yields a ``computational work'' of
$10^{36}$ kg yr.  The mass of the biosphere is liberally estimated as
$10^{20}$kg (the mass of layer of water 1 km thick covering the Earth) and its age is about one
billion years, so the ``computational work'' is $10^{29}$kg yr. In the comparison  of
``computational work,'' the Sun wins by 7 orders of magnitude, in the comparison of
``\depth'' the biosphere wins by 3 orders of magnitude.  The ``\depth'' of the whole
system is dominated by the biosphere but, if we divide the ``computational work'' of
simulating the whole solar system by its total  mass, we get back very nearly the 
``\depth'' of the Sun with a negligible contribution from the biosphere.

The point of this exercise is not the numbers themselves but to illustrate 
the fact that  \depth\  has the property
of {\em maximality}: for a system $AB$ composed of two independent subsystems, $A$ and $B$ with \depth\
$\D(A)$ and
$\D(B)$, respectively, the depth $\D(AB)$ of the whole is the maximum over the
subsystems, $\D(AB)=\max\{\D(A),\D(B)\}$.  For homogeneous equilibrium systems
with short range correlations \depth\ is nearly independent of system size, like {\em intensive}
properties in thermodynamics such as temperature or pressure.  The \depth\ of complex systems
may be dominated by a small, perhaps fractal, part of the whole system.
Parallel \depth\ has the property of maximality but logical \depth\ or other measures based on sequential computing do not.  

Having agreed that parallel rather than sequential time is a better choice for defining \depth\ 
we need to decide between local communication (embodied in the CA) and non-local communication (embodied in the PRAM).  Though the passage of time is required for the emergence of complexity, it does not seem likely that simply moving information from one place to another increases complexity.  It is the interaction of information embodied in logical operations that leads to novelty and, potentially, complexity.  Furthermore, allowing non-local communication creates a category of simple processes that can be simulated in parallel time that scales as the logarithm of the sequential time.  For example,  simulating the trajectory of a particle diffusing for  time $t$ can be carried out in $\ord(\log t)$ steps on a PRAM but requires $\ord(t)$ steps on a CA.  As we shall see below, similar speed-ups hold for a number of simple physical processes simulated on a PRAM.  The central proposal of this paper is that PRAM time, which counts logical steps while discounting both communication and hardware, is the computational resource that is best correlated with the potential for generating physical complexity,   

How many processors should the PRAM be allowed? 
Following the usual definitions in parallel complexity  theory we allow number of processors
to be very large, specifically, for a system with $N$ degrees of freedom, the number of
processors is bounded by a polynomial in
$N$.  
In most cases, the number of processors that we are envisioning is much larger than is
technologically feasible but the aim here is a fundamental measure of the number of logical steps
needed to carry out a simulation, not a practical method of actually doing so.

Alternatives to polynomial bounds on the number of processors might seem natural.  A point of
view with physical appeal is that the number of processors should
be proportional to the number of relevant degrees of freedom of the system. 
Another possibility is that the number of processors should be exponentially 
bounded.  Neither of these
extremes yields a useful or robust measure of parallel time.  Exponential parallelism allows all possibilities to be explored and evaluated simultaneously.  Any
problem that can be solved in polynomial time with polynomially many processors can be solved in constant parallel time with exponentially many processors.   Allowing exponentially many processors
eliminates the distinction between long and short histories and even very complex systems would be seen to have short histories.  On the other hand, insisting that the
simulation uses no more processors than there are degrees of freedom may rule out the use of efficient parallel algorithms that require more than linearly many  
processors.  For example, finding minimum weight paths between pairs of vertices on an arbitrary graph with positive weights on edges can be carried out in polylogarithmic time with more than linear processors but there is no known algorithm that runs in polylogarithmic time with linear processors.  The most fruitful definition of  parallelism in theoretical computer science
permits polynomially many processors and yields a sharp distinction between a class of problems
that can be solved quickly in parallel and those that are inherently sequential.  Whether this is
the best definition for distinguishing long from short natural histories can only be settled by
examining the proposal in various contexts.  A key hypothesis of this work is that the polynomial hardware bound that has proved the most fruitful in theoretical computer science is also the correct choice for an irreducible measure of history for natural systems.

\Depth\ is defined for statistical ensembles of system states rather than for individual system states.  In this regard, \depth\ is similar to other quantities defined in  statistical physics such as entropy. Statistical physics is a general framework for the study of complex natural systems and is a source of numerous tractable examples where \depth\ can be studied and related to various
physical properties or manifestations of complexity.   The \depth\ of a system refers 
to the average parallel running time of a Monte Carlo simulation that generates a typical
state of the system.  Monte  Carlo simulations are, loosely speaking, simulations that use random numbers.  In the analysis of \depth\, Monte Carlo simulations are used to sample from a distribution of system states. In practical Monte Carlo simulations~\cite{NeBa99} pseudorandom numbers 
are used but for the present purpose we employ a model of computation 
that is equipped with a supply of true randomness. 

The statistical framework serves to highlight the crucial
role  played by randomness in the evolution of complex systems.  Randomness  arises from initial conditions, external perturbations or, most  fundamentally, from quantum de-coherence.  A
generic feature of complex  histories is that some random choices are ``frozen--in" and produce 
a lasting effect on the system.  In Darwinian evolution, differential 
reproduction freezes in favorable random mutations.  Some features  
of all living organisms, such as the machinery of DNA 
replication and protein synthesis, were set up early in the history of 
life and have been highly conserved for hundreds of millions of years.  
Presumably, some  details of this basic machinery are 
arbitrary and could  work as well in other ways while other features 
are  uniquely determined but had to be 
discovered by random exploration of possibilities. Randomness plays a role in complex systems both in finding unique solutions to problems and choosing among feasible alternatives.

A fascinating question, debated in semi-popular expositions \cite{Gould, Morris} but not yet accessible to scientific study, is the extent to which the history of the Earth would repeat itself if played over many times.  Would life always arise and, if it did, would it always involve a central, information bearing, biopolymer, and, if so, would it always be DNA or something very similar?  The simulations implied by a measurement of the \depth\ of the biosphere, if they could be carried out, would allow one to answer these questions, as would observations of many Earth-like systems orbiting Sun-like stars.  The \depth\ of the biosphere is the running time of a simulation of a generic Earth, conditioned to the subset of runs that yield life.  It is an assumption of the whole set-up that such a simulation would produce life with reasonable frequency.  If, on the contrary, almost all runs of the Earth simulation are barren then it would not make sense to speak of the \depth\ of the biosphere.

Theoretically tractable examples of the freezing--in of random choices can be
found in statistical physics.  One particularly illuminating model is diffusion limited
aggregation (DLA)~\cite{WiSa}.  Diffusion limited aggregation generates fractal patterns like the
one shown in Fig.\ \ref{fig:four_fractals}(d).  It serves as a useful model for a variety of physical processes including
electrodeposition and fluid flow in porous media.  The dynamical rules for DLA are very
simple:  the initial condition is a single particle fixed at the origin.  A second particle is
released far from the origin and moves randomly until it either touches the particle at the
origin or drifts too far away from the origin.  If the moving particle reaches 
the fixed particle at the origin it
sticks and the aggregate now consists of two particles.  If the moving particle drifts far away it is considered
lost from the system.  In either case, after the first particle is disposed of, a second particle is
released and randomly walks until it is lost or sticks to the existing aggregate.  Particles are
successively released far from the aggregate and move as a random walk until they stick to
the aggregate or are lost.  What is clear from Fig.\ \ref{fig:four_fractals}(d) is that the location of major branches is
determined by random accidents occurring early in the growth.  In DLA random 
choices influence later growth for the simple reason
that the outermost tips of the aggregate are much more likely to grow than the inner recesses
so that branches which  already extend farthest from the origin grow the fastest which then
makes them successful in later epochs.  As we shall see, DLA dynamics leads to a long history and substantial \depth.

\section{Computational complexity and parallel computing}
\label{sec:cc}
Computational complexity theory determines the scaling of computational resources needed to
solve problems as a function of the size of the problem~\cite{Papa,GrHoRu}.  Although the theory
can be formulated with respect to various models of computation and is motivated by questions
raised by real computational problems, computational complexity theory is fundamentally about
the logical structure of problems.  This abstract view is most clearly manifest in ``descriptive
complexity theory''~\cite{Im99} where computational complexity is defined in terms of the
size and structure of formal logical expressions describing a problem.  Our interest is in  the
logical structure of dynamical processes occurring in the physical world rather than practical
questions about how to best simulate these processes, nonetheless, it is easier to think about
computational complexity theory in terms of two concrete models of parallel computation--the
PRAM and families of Boolean circuits.   A PRAM consists of  a number of simple processors
(random access machines or RAMs) all connected to a global memory, as shown in Fig.\ \ref{fig:pram}.  Although a RAM is defined
with much less computational power than a real microprocessor such as a Pentium, it would not
change the general arguments presented here to think of a PRAM as being composed of many
microprocessors all connected to the same random access memory.  PRAM processors run
synchronously and each processor runs the same program but processors have distinct integer
labels and thus may follow different computational paths.  During each parallel
step, processors carry out independent actions;  communication between processors occurs from
one step to the next via reading and writing to memory.  Since all processors access the same
global memory, provision must be made for handling memory conflicts.  One possibility is the priority
concurrent read, concurrent write (CRCW) PRAM in which many processors may attempt to write
to or read from the same memory cell at one time.  If there are conflicts between what is to be
written, the processor with the smallest label succeeds in writing.

The two most important resources associated with PRAM computation are parallel time, $T$,
and number processors or hardware, $H$. The objective of computational complexity theory is to
determine how parallel time and number of processors scale as a function of the size of the
problem and to study the trade-off between them. Another resource is non-uniformity, which is the amount of auxiliary information, such as externally supplied constants, needed to carry out a computation.  In what follows we consider only uniform circuits or PRAM programs.  Monte Carlo simulations require an additional resource--random numbers.  We suppose these values are available in special memory cells.  

\begin{figure}
\includegraphics[width=5in]{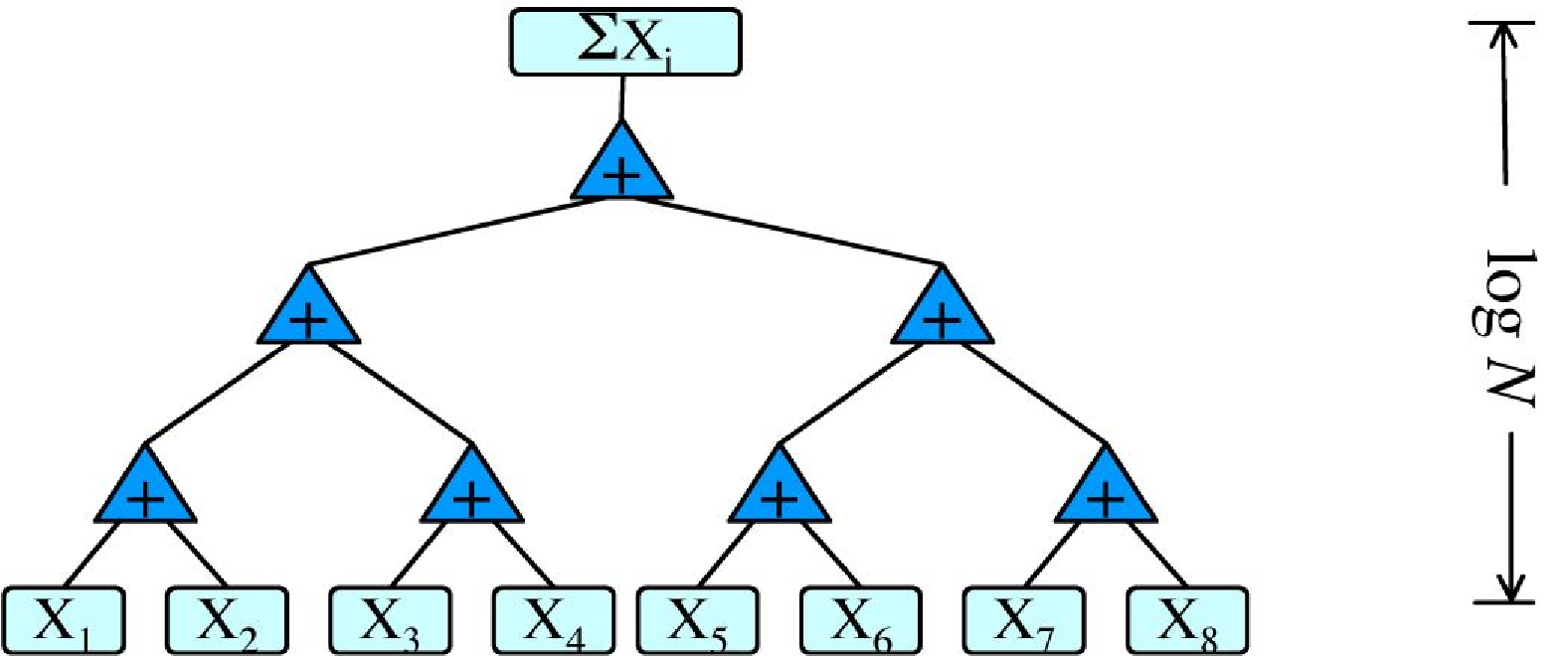}
\caption{Addition of $N$ numbers in $\log N$ parallel steps.}
\label{fig:tree}
\end{figure}

A problem that can be solved by $H$ processors running for $T$ steps could also be solved by
one processor running for $HT$ steps since the single processor can, in sequence, carry out the
work of the $H$ processors. On the other hand, it is not obvious whether the work of a single
processor can be re-organized so that it can be accomplished in substantially less time by many processors.  Several examples will help illustrate this point.  The first is adding
$N$ numbers.  On a PRAM with $N/2$ processors this can be done in $\ord(\log N)$ time
using a binary tree as shown in Fig.\ \ref{fig:tree}.  In the first step, processor one adds the first and second numbers and puts the result in memory, processor two adds the third  and fourth numbers and so on.  After this first parallel step is completed, there are $N/2$ new numbers to add and these are again summed in a pairwise fashion by $N/4$ processors.  The summation is completed after $\ord(\log N)$ steps.  This technique is rather general and applies to any binary, associative operation.  It is clear that the same method could be used to generate random walk trajectories quickly in parallel.     Some more difficult tasks can also be carried out quickly in parallel.  The problem of identifying minimum weight paths between pairs of points on a weighted graph is relevant to the discussion of a number of physically motivated models.  Given a graph with $N$ nodes and vertices with positive weights, shortest paths between all pairs of nodes can be identified in $\ord(\log^2 N)$ parallel time on a PRAM using $N^3$ processors.  Both addition and minimum weight paths have {\em efficient} parallel solutions in the sense that they can be solved on a PRAM with {\em polynomially} many processors in {\em polylog} time.   ``Polynomial" means bounded by some power of the problem size and ``polylog'' means bounded by some power of the logarithm of the problem size.  

\begin{figure}
\includegraphics[width=4in]{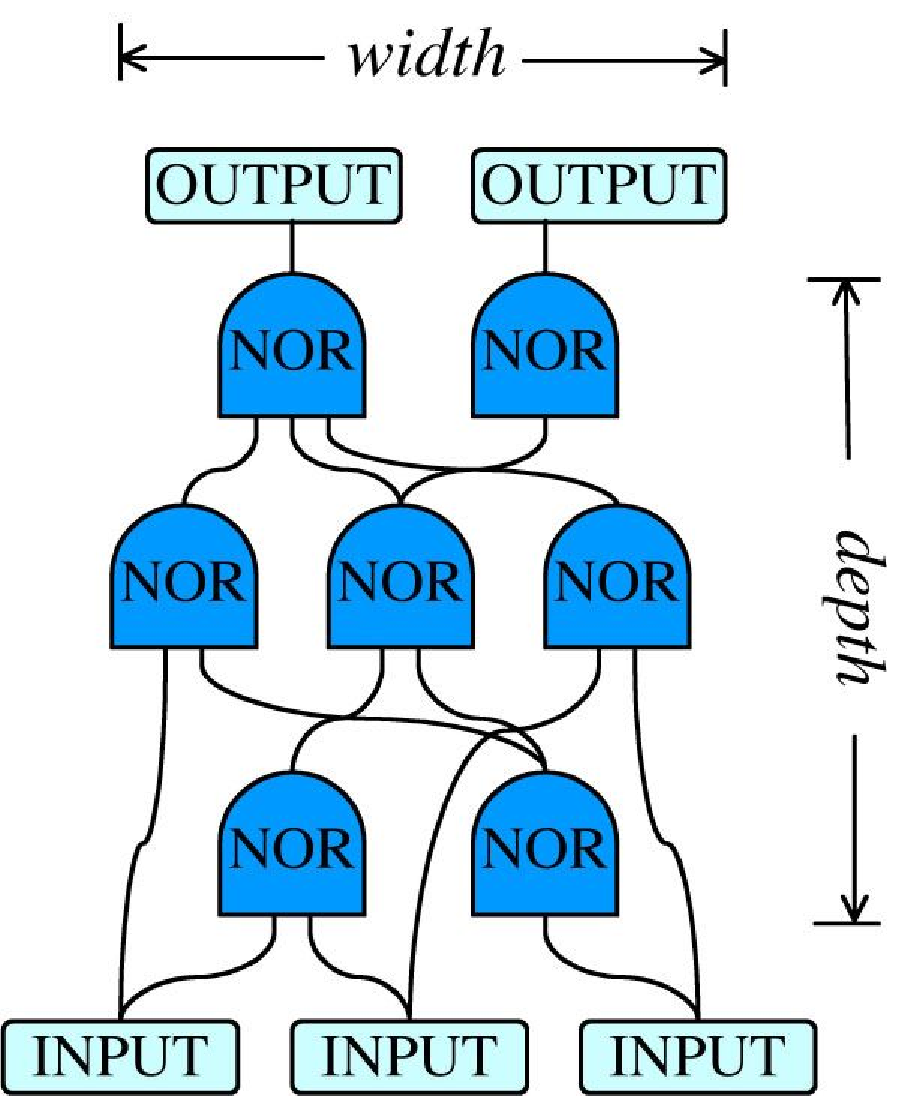}
\caption{A Boolean circuit.}
\label{fig:circuit}
\end{figure}

While many problems  have efficient parallel solutions, it is thought that there exists problems that can be solved in polynomial time by a single processor but have no efficient parallel solution.  Problems of this kind are called {\em inherently sequential}.  An example of a problem that is believed to be inherently sequential is evaluating the output of a Boolean circuit with given  inputs.  A Boolean circuit is composed of logic gates connected by wires.  The
gates are arranged in levels so that gates in one level take their inputs only from gates of the
lower levels so that there is no
feedback in the circuit.  At bottom level of the circuit are TRUE or FALSE inputs and at the top
level are one or more outputs.  Circuits can be classified by their size.  The depth of a circuit is
the number of levels and the width is the largest number of gates in a level.  As we shall see, the
notion of \depth\ for physical systems is closely related to circuit depth.  Figure \ref{fig:circuit} shows a Boolean circuit composed of NOR gates, which can be used by themselves to construct an arbitrary circuit. 

Given some
concisely encoded description of the circuit and its inputs, the problem of obtaining the outputs is
known as the {\em circuit value problem} (CVP).  Clearly one can solve CVP on a PRAM in parallel
time that is proportional to the depth of the circuit since each level of the circuit, starting from
the bottom level and working up, can be evaluated in a single parallel step.  On the other hand,
there is no known general procedure for speeding up the evaluation of a Boolean circuit to
polylog parallel time and it is presumed that there is none.  The logical structure of adding
$N$ numbers is sufficiently simple that a wholesale substitution of
hardware for time is possible whereas for CVP the logical structure is arbitrary and there is no known general procedure for reducing the depth of the problem by using (polynomially) more hardware.

At the present time, there is no proof that CVP cannot be solved efficiently in parallel.  The best one can do is show that CVP is \PP-complete.  To
understand the meaning of \PP-completeness, we must first introduce the complexity classes
\PP\ and \NC\ and the notion of {\em reduction}.  \PP\ consists of the class of {\em decision} problems that
can be solved by one processor in polynomial time and 
\NC\ consists of the class of decision problems that can be solved in polylog parallel time on a PRAM with polynomially many processors. A decision problem is a problem with a `yes' or `no' answer.  Clearly $\NC \subseteq \PP$ but it is not known whether the
inclusion is strict.  A problem
${\cal A}$ is reduced to a problem
${\cal B}$ if a PRAM with an {\em oracle} for ${\cal B}$ can be used to solve ${\cal A}$ in polylog
time with polynomially many processors.  An oracle for ${\cal B}$ is able to supply a solution to
an instance of ${\cal B}$ in a single time step.  Intuitively, if ${\cal A}$ can be reduced to ${\cal
B}$ then ${\cal A}$ is no harder to solve in parallel than ${\cal B}$.

A problem in \PP\ is \PP-complete if all other problems in \PP\ can be reduced to it.  CVP is an
example of a \PP-complete problem.  It follows from the definition of reductions that if any
\PP-complete problem is in \NC\ then $\PP=\NC$ and all problems that can be solved in
polynomial time by one processor can be solved in parallel in polylog time with
polynomial hardware.  However, no one has found a fast
parallel algorithm for any \PP-complete problem and it  is generally assumed 
that, in fact,  $\PP
\neq \NC$ from which it follows that \PP-complete problems
cannot be solved in polylog time with polynomial hardware.   The hypothesis that some
problems are  inherently sequential and cannot be solved in a small number of parallel
steps is crucial to the ideas developed here, without this hypothesis no physical system would
have much \depth.

The proof that CVP is \PP-complete proceeds by showing that any  Turing
computation can be mapped onto the evaluation of a Boolean circuit.  The  proof follows from the recognition that Boolean circuits are themselves a universal model of
computation.  Since it is hardwired, a Boolean circuit is equivalent to a PRAM running a
specific program for specific problem size.  A {\em family of Boolean circuits}, with one circuit
for each problem size is equivalent to a PRAM running a specific program designed to solve a
problem of arbitrary size. The resources of circuit depth and circuit width are nearly identical
to the PRAM resources of parallel time and number of processors, respectively.   For example,
\NC\ can be defined as the class of problems solved by uniform families of Boolean circuits whose
width is polynomial in the problem size and whose depth is polylogarithmic in
the problem size.  For circuit families, uniformity is the requirement that the design of successive members of the family can be easily computed.  We shall only consider uniform circuit families.  Families of Boolean circuits provide a useful alternative to PRAMs in
thinking about the physical notion of \depth.  A Boolean circuit equipped with random inputs
can perform a Monte Carlo simulation of a physical system and physical \depth\ corresponds to
the circuit depth of the optimal Boolean circuit that performs the simulation.

The physical models discussed here are mainly associated with the complexity classes \PP\
and \NC, however more inclusive complexity classes requiring more computational work are also related to physical problems. Two classes of particular interest are \NP\ and \PS~\cite{GaJo}.  In terms of Boolean circuit
families,
\NP\ consists of problems that can be solved by monotone semi-unbounded fan-in circuit families with logarithmic depth and exponential width and \PS\ consists of problems that can be solved by circuits with polynomial depth and exponential width~\cite{Ve}.  Semi-unbounded fan-in circuits allow unbounded fan-in for OR gates and but constant fan-in for AND gates.  Monotone circuits have no NOT gates except at inputs.  \PS, which stands for polynomial space, can also be defined as the set of problems that can be solved by a Turing machine with polynomially  bounded memory.  It is not difficult to see that $\PP \subseteq \NP \subseteq \PS$ but  there is no proof that either inclusion is strict.  Reductions and completeness can be defined yielding notions of \NP-completeness and \PS-completeness.  Under the well-accepted hypotheses that $\PP
\neq \NP$,
both \NP-complete and \PS-complete problems require more than polynomial computational work for their solution. 

\NP-complete problems are typically related to optimization problems, such as the famous Traveling Salesman
Problem, and finding the best solution requires an exhaustive search among possibilities.   Another way of seeing the difference between  \PP\ and \NP\ is to compare the \PP-complete circuit value problem with its \NP-complete analog, satisfiability (SAT).  CVP asks whether a Boolean circuit with given inputs evaluates to TRUE.   Satisfiability problems ask whether {\em there exists} a set of inputs for which a circuit evaluates to true.  

A physically motivated \NP-complete problem is to find the ground state energy of an Ising spin glass.  It is plausible to suppose that Nature is not able to solve \NP-hard problems more efficiently than a computer so that physical spin glass
systems are almost never found in their ground state.  On the other hand, though it is a very difficult problem to find ground states of spin glasses or optimum tours for traveling salespersons, the ground state or tour is itself not complex in the  intuitive or physical sense of the word.  

Many \PS-complete problems are stated in terms of who wins a two-player board game given ideal play by both players.  A problem of this kind, involving players A and B can be stated in the form, ``Does there exist an opening move for A such that for all possible  first moves for B there exists a second move for A such that for all possible second moves for  B  $\ldots$ there exist a winning move for A?"  Note that this kind of problem is characterized by a long string of quantifiers alternating between ``there exists" and ``for all."  The alternation of quantifiers can also be seen in Boolean versions of  \PS-complete problems.  The \PS-complete analog of circuit value or satisfiability problems is quantified Boolean formulas (QBF), which asks whether the quantified 
Boolean formula  $Qx_1 Qx_2\ldots Qx_n F(x_1, x_2, \ldots , x_n)$ is TRUE.  Here $Q$ is a quantifier, either ``there exists" ($\exists$) or ``for all" ($\forall$) and $F$ is a Boolean formula over logical variables $x_1, x_2, \ldots , x_n$. Satisfiability has only $\exists$ quantifiers and CVP has no quantifiers, only specified variables.

The most well-developed area of computational complexity theory is the study of worst-case
decision problems.  For example, \PP\ consists
of the class of decision problems for which a polynomial time bound holds for all instances of the
problem. Our interest is in the average-case complexity of sampling distributions--\depth\ is the
average parallel time needed to generate a sample from a distribution of physical
states.  Questions of this kind are much less well-understood~\cite{JeVaVa,Jerrum03, BeChGoLu}.  We can put upper bounds on the complexity of sampling by explicitly analyzing the most efficient known parallel
sampling algorithm. Lower bounds are much more difficult to obtain and theoretical tools for directly
tackling such problems are not yet available.

\section{Statistical physics}
\label{sec:sp}

Statistical physics~\cite{PlBe, Ka00} is the branch of physics dealing with emergent behavior in systems
having many interacting components.  The field was originally developed to provide a
microscopic underpinning to the sciences of thermodynamics and
hydrodynamics and to give explicit tools for calculating the undetermined constants and
functions appearing in these macroscopic theories.  For example, thermodynamics allows one
to compute a variety of properties of a liquid such as its compressibility or heat capacity once the free energy is known as a function of temperature and density.   Statistical physics supplies a framework for
calculating the free energy directly from the microscopic interaction of the constituent
molecules.  Similarly, hydrodynamics allows one to predict the time evolution of a flowing
liquid once certain macroscopic properties of the liquid such as its viscosity are known and
statistical physics provides ways of computing transport coefficients such as viscosity directly
from the microscopic interaction of the molecules.  Compressibility, heat capacity and viscosity are not defined for individual molecules, they are emergent properties of large assemblies of molecules.   The purview of modern statistical
physics has broadened to include increasingly complex phenomena. Statistical physicists are turning their attention to complex biological materials and to the analysis of  models intended to characterize aspects of very
complex phenomena such as occur in macroevolution, finance or the growth of the Internet.

Statistical physics, as its name implies, gives a probabilistic description of nature and the
fundamental objects of the theory are probability distributions of system states or system
histories.  For systems in thermodynamic equilibrium at absolute temperature $T$, the probability,
$P[\sigma]$ of finding state $\sigma$ is given by the Gibbs distribution,
\begin{equation}
\label{eq:gibbs}
P[\sigma]= \frac{1}{{\cal Z}}\exp(-H[\sigma]/k_B T)
\end{equation}
where $H[\sigma]$ is the energy of the state, $k_B$ is
Boltzmann's constant, and ${\cal Z}=\sum_\sigma \exp(-H[\sigma]/k_B T)$ is the required
normalization, known as the ``partition function."  The state $\sigma$ is some concise, natural specification of the microscopic degrees of freedom. For example, the state of a classical gas is a list of the positions and velocities of the constituent molecules.

For systems out of equilibrium there is typically no closed form expression for
the probability distribution and, instead, probabilities are implicitly defined by the
stochastic dynamics of the system.  DLA is an example of a system where the ensemble is defined by the dynamics of the systems.  

Although statistical physics is a probabilistic theory, its
conclusions often apply to individual systems.  For example, in an equilibrium sample of gas with $N$ particles the
average energy of the ensemble will be proportional to $N$ but the fluctuations in energy from
one sample to another will be proportional to $\sqrt N$ so that the average energy is a very good
estimate of the energy of an individual system if $N$ is of the order of Avogadro's number.  The energy is said to be ``self-averaging" since its value in one system is nearly the same as the ensemble average. Even in equilibrium systems,  the situation can be more complicated because of the presence of multiple thermodynamic phases with different macroscopic properties.  At its triple point, water may be in liquid, vapor or solid form, each with very different properties.  If there are several coexisting phases the ensemble
is partitioned into several components having small fluctuations within each component but large
differences from one component to the next.  Averaging must be done over a single
phase rather than over the whole ensemble to obtain results that apply to an individual system in that
phase.  For equilibrium systems, the Gibbs phase rule limits the number of coexisting phases but for non-equilibrium systems the phase structure may be very complicated.   

Like computational complexity theory, statistical physics is a scaling theory whose most robust results apply to systems that can be uniformly scaled up to having $N$ degrees of freedom with $N$ large.  The behavior of various observables as a function of $N$ and the asymptotic properties as $N \rightarrow \infty$ are the primary concerns of the theory.   

\section{The definition of \Depth\ in Statistical Physics}
\label{sec:def}

 Following the usual approach of
both statistical physics and computational complexity theory, consider a sequence of similar systems with increasing numbers of degrees of freedom, $N$.   Let ${\cal A}$ refer to a family of systems of increasing size and let $A$ be one member of the family of size $N$.  $A$ refers to a concise description of the degrees of freedom or microstates for the system together with a probability distribution for those degrees of freedom.  As discussed above, the probability distribution may be specified by a closed form expression like the Gibbs distribution or it may be specified to be the result of a stochastic dynamical process.  The objective is to {\em sample} the probability distribution of the microstates of the system using a
PRAM (or, alternatively, a uniform family of Boolean circuits) with polynomially bounded hardware equipped with a supply of random
bits.  Here ``sample" means to generate a microstate from the ensemble with the correct probability.  Monte Carlo algorithms in statistical physics typically sample distributions. The most efficient feasible algorithm for sampling microstates is defined to be the Monte Carlo algorithm whose parallel time on a PRAM with polynomially bounded hardware is
asymptotically smallest in the limit of large $N$.    Here, then, is a working definition of \depth:
\begin{quote}
{\em The physical  \depth\, $\D(A)$ of a system $A$ is the average parallel time needed to generate a typical system state using the most efficient, feasible Monte Carlo algorithm for ${\cal A}$.}
\end{quote}
Equivalently,  physical \depth\ is
the circuit depth of a Boolean circuit with random inputs that simulates typical states of a
system. The width of the Boolean circuit is bounded by a polynomial in the number of degrees of
freedom of the system.

$\D(A)$ is the {\em absolute} \depth\ of $A$, meaning that states of $A$ are generated from only a short program and random bits.  One can also consider {\em relative} \depth.  Suppose that typical states of some other system $B$ are available and we want to know how much additional \depth\ is associated with generating states of $A$ given states of $B$.  Let the relative \depth\  of $A$ given $B$, $\D(A|B)$ be defined in the same way as absolute \depth\ except that, in addition to random bits, the computer sampling $A$ also has free access to typical states of $B$.

Several properties of  absolute  and relative \depth\ follow directly from the definitions.  First, suppose that $A$ and $B$ are two independent systems.  Let $A  B$ be the system composed of independent subsystems $A$ and $B$ (described by the product measure), then, 
\begin{equation}
\D(A B) = \max\{\D(A),\D(B)\}.
\end{equation}
This maximal property follows from the observations that the most efficient feasible simulation of $A B$ is obtained by running the most efficient feasible simulations of $A$ and $B$ independently in parallel.  The overall parallel time is simply the maximum of the running times of the two simulations.  

Two straightforward inequalities hold for relative \depth.  Since one can always choose to ignore states of $B$ in sampling $A$, relative \depth\ is no greater than absolute \depth.  On the other hand, if the steps required to sample $B$ are counted and then added to the relative \depth\ of $A$ given $B$ the results must not be less than the absolute \depth\ of $A$ since otherwise the algorithm associated with the absolute \depth\ of $A$ would not be the most efficient--it would be better to generate states of $B$ and then use them to generate states of $A$.   Thus, for any $A$ and any $B$, we have the inequalities,
\begin{equation}
\D(A|B)  \leq \D(A) \leq \D(A|B) + \D(B).
\end{equation}

The above definition of \depth\ requires that the distribution of system states is exactly sampled.  In some cases it may be appropriate to define \depth\ with respect to approximate rather than exact sampling.  The need for approximate sampling is highlighted by the case of pseudorandomness.   A long string of bits is pseudorandom if it is generated by a deterministic computation from a much shorter random bit string (the seed) and the distribution of the long strings is sufficiently close to the uniform distribution.  Pseudorandomness is used in Monte Carlo simulations and in cryptography~\cite{goldreich01}. The criterion for being sufficiently close to the uniform distribution depends on the application in question.  How should we define the \depth\ of a pseudorandom bit string?  On the one hand, considerable parallel time may be needed to generate a pseudorandom string from a seed suggesting that pseudorandom strings have considerable \depth.  On the other hand, a  good pseudorandom string, like a truly random string, has no intuitive complexity and should be assigned no \depth.  A resolution of this conflict is to define \depth\ as the minimum parallel time for sampling a distribution that is sufficiently close to the real distribution.   Pseudorandom distributions are then adequately simulated by the uniform distribution and assigned no \depth.  The meaning of two distributions being sufficiently close is an open question that may depend on the type of system studied.  A proposal for defining and measuring the \depth\ of deterministic chaotic maps using approximate sampling is discussed in Ref.\ \cite{Mac02}.

\section{\Depth\ in Statistical Physics}
\label{sec:examples}

In this section we consider the \depth\ of some well-known systems in statistical physics.  We will examine several non-equilibrium growth models, including diffusion limited aggregation, and an equilibrium spin system, the Ising model.  These models are highly simplified representations of the real world, simple enough that we can develop an understanding of their \depth. Nonetheless they are rich enough to possess some of the properties that are intuitively associated with complexity.  

\begin{figure}
\includegraphics[width=5in]{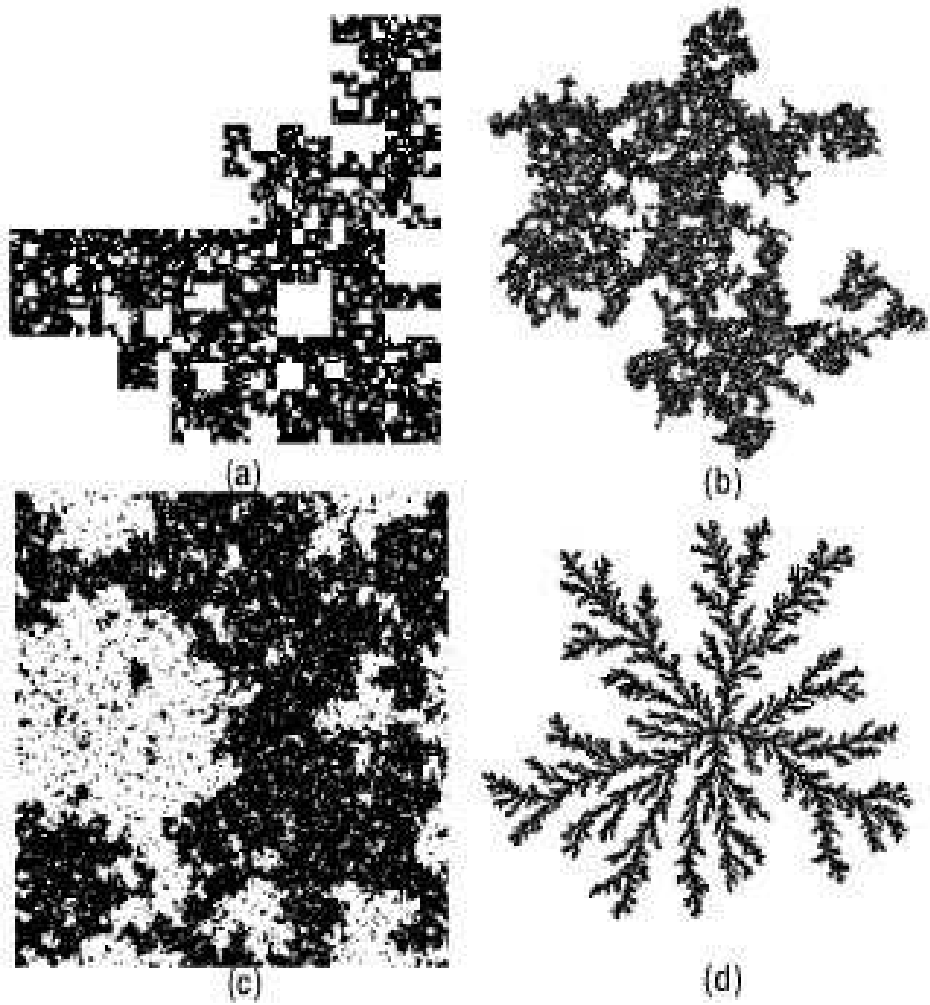}
\caption{Examples of two-dimensional fractal patterns: (a) Mandelbrot percolation, (b) invasion percolation (c) the critical Ising model and (d) diffusion limited aggregation.}
\label{fig:four_fractals}
\end{figure}

One intuition is that physical complexity is associated with long range spatial correlations.  Mandelbrot~\cite{Man83} first pointed out the ubiquity of self-similarity in the world and coined the term ``fractal" to describe self-similar structures.   Figure \ref{fig:four_fractals} shows fractal structures generated by four systems--Mandelbrot percolation, invasion percolation, the Ising model at criticality and diffusion limited aggregation.  We will introduce each of these systems and discuss the \depth\ of the states they generate. One of the surprising conclusions of this section is that fractal structures have widely varying   \depth.   Long range spatial correlations appear to be a necessary but not sufficient condition for \depth. 

\subsection{Equilibrium Ising model}
\label{sec:ising}

As originally conceived, the Ising model represents a magnetic material and its degrees of freedom represent ``electron spins'' on a lattice.  The system
state is specified by the values of the spin variables
$\sigma=\{s_j \}$ where spin, $s_j$ resides on the lattice site indexed by $j$ and can take on
one of two values,
$+1$ and $-1$, referred to as ``spin up" and ``spin down," respectively.   Spins interact locally
according to an energy function,
$H[\sigma]$,
\begin{equation}
\label{eq:isingenergy}
H[\sigma]=-\sum_{(i,j)} J_{ij} s_i s_j
\end{equation}
where the summation is over all pairs of nearest neighbor sites on the lattice and $J_{ij} \equiv 1$.  The energy
function expresses the preference for neighboring spins to align in the same
direction and it is easy to see that there are two possible minimum energy
states of the system, one with all spins up ($+1$) and the other with all spins down ($-1$). 
At low temperatures, the state of the system is near one or the other of these completely
ordered states.  The degree of order of a state can be measured by the
``magnetization,''
$m=(1/N)
\langle |
\sum_j s_j |\rangle$ where the angular brackets refer to an average over the
Gibbs distribution and $N$ is the number of spins.  At low temperature, $m$ is near one but 
as the temperature is raised, the disordering effect of thermal fluctuations becomes more
important.  At infinite temperature, Eqs.\
\ref{eq:gibbs} and
\ref{eq:isingenergy} imply that all states are equally likely, that is, each spin behaves as an
independent random variable and the magnetization vanishes as $N \rightarrow \infty$.  In the
large $N$ limit, there is a sharply defined phase transition at a temperature $T_c$
between a low temperature, ordered phase with nonzero magnetization and a high temperature,
disordered phase with zero magnetization. 

Non-trivial collective behavior in the Ising model is quantified by the correlation length.
Except at the critical point, the spin autocorrelation function, $\chi(r)=\langle s_j s_{j+r}
\rangle-\langle s_j
\rangle \langle s_{j+r} \rangle$ decays exponentially in the distance $r$ with a characteristic
length
$\xi$, called the {\em correlation length}.  $\chi(r)$ measures 
the likelihood that spins separated by $r$ fluctuate coherently.  The subtracted term  insures that $\chi(r)$ decays rapidly if the system is ordered.  As the critical point is
approached from either the high temperature or low temperature side, the correlation length
diverges as a power law in the distance from the criticality,
$\xi \sim |T-T_c|^{-\nu}$ where $\nu$ is known as the {\em correlation length exponent}.  In
the critical phase exactly at $T=T_c$ the correlation length is infinite and
$\chi(r)$ decays as a power law in $r$,  $\chi(r) \sim 1/r^{2+d-\eta}$ where $\eta$ is a second, independent, critical exponent.  At the critical point the correlation between spins decays slowly and fluctuations occur on all length scales.  In fluid systems, the presence of correlated fluctuations on the scale of the wavelength of visible light produces the stunning phenomenon of ``critical opalescence''--a fluid which is transparent away from its liquid-vapor critical point becomes milky white at the critical temperature and pressure.    Figure \ref{fig:four_fractals}c shows a critical configuration of the Ising model and reveals the long range correlations.  The large cluster spanning the system is a fractal object.

Critical points in physical systems such as fluids and magnets and models such as the Ising model have been the object of intensive study in statistical physics.  One of the striking features of critical phenomena is {\em universality}: the critical exponents and certain other properties are precisely the same within large classes of disparate systems all sharing the same underlying symmetry in the way they order.  For example, the liquid-vapor critical point, the critical point of uniaxial magnets and the critical point of the Ising model are all in the same universality class for which $\nu\approx0.63$ and $\eta\approx0.04$.  Universality is a consequence of the fact that long range behavior near a critical point depends only weakly on the details of the microscopic system.  The explanation of universality and techniques for calculating critical exponents can be found within the framework of the {\em renormalization group} methodology~\cite{BiDoFiNe}.

Equilibrium systems display their most complex behavior at critical points so it is here that we would expect 
\depth\ to be greatest. A hint that \depth\ might be greatest at the critical point comes from the experimental observation that the time to reach equilibrium becomes very long there.  This phenomena is known as {\em critical slowing} and is a direct result of the divergence of the correlation length.  The \depth\ of the equilibrium Ising model is determined by the running time of the most efficient parallel algorithm for sampling the  Gibbs distribution. It is easy to see that at zero
and infinite temperature, the \depth\ of Ising states is very low: at zero temperature a
single random bit directly determines every spin and at infinite temperature, each
spin is set by an independent random bit.  At both temperature extremes,
\depth\ reaches a minimum in conformance with the intuition  that completely ordered and
completely random systems lack complexity. Away from these extremes \depth\ increases but
since the most efficient parallel algorithm for the Ising model is not known we can only set upper
bounds.  

One of the most widely used algorithm for sampling equilibrium states is
the Metropolis algorithm~\cite{Met, NeBa99}.  The Metropolis algorithm equips the Ising model with stochastic dynamics  and is guaranteed to approach equilibrium when run long enough.  Each elementary move of
the Metropolis algorithm consists of choosing a spin and proposing to  ``flip'' it from its current
state to minus the current state.  If the energy is lowered, the flip is actually
carried out.  If the energy is raised by $\Delta H$, the flip is carried out with
probability
$\exp(-\Delta H/k_BT)$, otherwise the spin is left unchanged.  One {\em sweep} of the
Metropolis algorithm consists of attempting to flip each spin once. It is easy to show that the
Metropolis algorithm will generate equilibrium states after sufficiently many sweeps but
more difficult to determine the {\em equilibration time}, the average number of sweeps
actually needed to  reach equilibrium.  The equilibration time for Metropolis dynamics for spin system is reasonably well-understood from a variety of theoretical and numerical studies.  A
single sweep of the Metropolis algorithm can be carried out in constant parallel time on a PRAM
by assigning one processor to each spin.  The Metropolis equilibration time is an upper
bound on the \depth\ of the Ising model. 

The Metropolis equilibration time is nearly independent of system size  above the
critical temperature~\cite{DySiViWe04}.  As the critical point is approached, the equilibration time diverges
roughly as the square of the correlation length and saturates when the correlation length
reaches the linear size of the system, $L$.  In the critical region for finite systems when $\xi
\geq L$, the equilibration time diverges as a power of the systems size $L^{z_{\rm M}}$ where
$z_{\rm M}$ is the {\em dynamic exponent} for Metropolis dynamics, $z_{\rm M} \approx 2$.  
The divergence of the equilibration time near critical points is the numerical manifestation of critical slowing
down seen experimentally.  In both cases, the long equilibration time results from the need for information to propagate across the system diffusively to set up spatial correlations the size of the system. Diffusion over a length scale $L$ requires time $L^2$.   

Metropolis dynamics is physical in the sense that
updating a given spin involves only information about neighboring spins.  Various experimental systems in the broad
``dynamic universality class'' of the Ising model with local dynamics and no conservation law for the order parameter also experience critical
slowing down in their equilibration time that is described by the dynamic
exponent, $z_{\rm M}$.   For Ising universality classes in two and three dimensions with a non-conserved order parameter the critical dynamic exponents are, $z^{(2)}_{\rm M} = 2.17$~\cite{NiBl00} and $z^{(3)}_{\rm M} = 2.04$~\cite{JaMaScZh99}, respectively.

Though physically realistic, the Metropolis algorithm is not the most
efficient method for simulating the critical region  of the Ising model. Cluster
algorithms~\cite{NeBa99,SwWa,Wolff,Swe} have less critical slowing and are the method of choice for
simulating many equilibrium critical phenomena.  These algorithms approach the equilibrium
state via a non-local process that flips large clusters of spins in a single step avoiding the need to diffuse information over large distances. Furthermore, the clusters that these algorithms identify are commensurate with the naturally occurring critical fluctuations in the system. As a
result, the dynamic exponent is much smaller than for local dynamics.  For the
three-dimensional Ising model the Swendsen-Wang~\cite{SwWa} cluster algorithm has dynamic exponent
$z_{\rm SW} \approx 0.5$~\cite{CoBa, OsSo04}. The performance of the Swendsen-Wang algorithm is better for the two-dimensional Ising model and may even be polylog in $L$ rather than polynomial.  Each
sweep of the Swendsen-Wang algorithm requires the identification of connected clusters of spins
which can be carried out in polylog time on a PRAM using standard methods for identifying
connected components~\cite{GiRy}.  Thus, at least in the critical region, an upper bound on the
\depth\ of the Ising model is set by the Swendsen-Wang algorithm.  This upper bound still diverges with system size but it is now seen to be controlled by $z_{\rm SW}$, which is much less than the ``physical'' value, $z_{\rm M}$. Another cluster algorithm introduced by Wolff~\cite{Wolff}  appears to have an even smaller dynamic exponent but is less well understood and more difficult to fully parallelize~\cite{MaCh03}.

\begin{figure}
\includegraphics[width=5in]{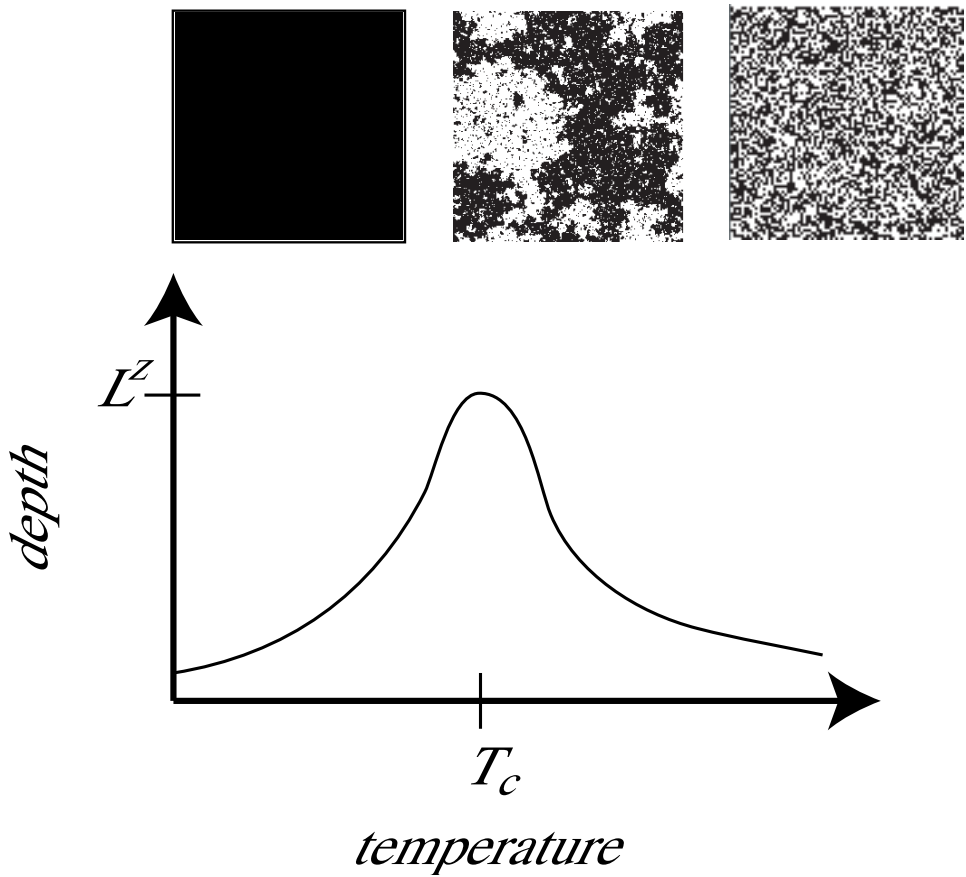}
\caption{(Bottom) Schematic plot of the \depth\ of the Ising model as a function of temperature for a system of size $L$. (Top) Typical states at zero temperature (left), infinite temperature (right) and the critical temperature (center).}
\label{fig:ising_depth_states}
\end{figure}

Upper bounds on the \depth\ of the Ising model and other equilibrium systems are
established by measuring the equilibration time of known parallel
algorithms. (Although we do not actually have PRAMs available to check the running
time of parallel algorithms, one can simulate a PRAM on a real
computer and extract the parallel running time that would have been required on a PRAM.) 
Establishing lower bounds is very difficult and one cannot rule out the possibility that there is
a yet  better way to simulate the Ising critical point.  Nonetheless, it seems safe to conclude from the behavior of the known algorithms that the \depth\ of the Ising model as a function of temperature for a system of size $L$ looks something like the sketch in Fig.\ \ref{fig:ising_depth_states}.  Typical Ising states at low, high and critical temperatures are shown above the graph.  The value of the dynamic exponent $z$ that describes how \depth\ increases with system size at criticality is unknown but bounded above by $z_{\rm SW}$.

A central ingredient of all cluster algorithms is the identification of connected components of a
graph. This problem can be solved in polylog time only if
$N^{3}$ processors are used~\cite{GiRy}, illustrating the distinction between  polynomial and linear
bounds on PRAM hardware.  If \depth\  were defined with a linear
bound on the number of processor, the Swendsen-Wang algorithms would require $L^{z_{\rm
SW}+1}$ parallel steps and \depth\ would not agree with the usual way of quantifying the performance of cluster algorithms.  By allowing more than linear hardware we discount the work involved in identifying connected components and correctly view cluster 
algorithms as providing a very short path from random numbers to critical spin configurations.

So far we have bounded the \depth\ of the Ising model with the number of Monte Carlo sweeps of either the Metropolis or
Swendsen-Wang algorithm.  Is it possible that parallelism would permit many sweeps to
be carried out in a much smaller number of parallel steps thus reducing the bound on \depth? The prospect for achieving reductions to polylog parallel time  by compressing many
Monte Carlo sweeps into a much smaller number of parallel steps is ruled out,
modulo accepting \PP$\neq$\NC, by the \PP-completeness proofs for natural decision problems associated with Metropolis and Swendsen-Wang dynamics~\cite{MaGr96}.  A \PP-completeness result even holds for zero temperature single spin flip dynamics~\cite{Moore97}.

In summary, the \depth\ of the Ising model is lowest at the extremes of high and
low temperature and increases as the critical point is approached 
where it apparently diverges as a small power of the system size though the exact
nature of that divergence is not known.  The fact that the \depth\ of the Ising model is greatest
at the critical point  agrees with the intuition that the equilibrium systems reach their greatest
complexity at critical points. 

\subsection{Mandelbrot percolation}
\label{sec:manperc}
Mandelbrot percolation~\cite{Man83, ChChDu} is an example of a ``hand-made'' random fractal that has been used as a simplified model of a porous material~\cite{Mac91a}.  A Mandelbrot percolation pattern is shown in Fig.\  \ref{fig:four_fractals}(a).  The construction of the pattern is a multiscale process.  To generate a two-dimensional Mandelbrot percolation pattern, the starting point is a black square of unit size.  The square is divided into four equal smaller squares, which are randomly whited out with probability $f$.  Each successive scale is created by subdividing squares of the previous scale into four smaller squares.  At each scale a fraction $f$ of the squares are randomly whited out. It is easy to see that the fractal dimension, $d_f$ of the resulting pattern is $d_f=2-\log(f)/\log(2)$. All of the decisions made during the construction of the pattern are independent and can be made simultaneously so there is evidently no history recorded in the final pattern.   A Boolean circuit that generates Mandelbrot percolation patterns is straightforward to design~\cite{MaGr96}.  Each square or pixel at the smallest scale is black only if it is not whited out at any level of the construction.  This condition is the logical NOR of all the random bits controlling the squares containing the pixel in question. Given arbitrary fan-in NOR gates, the circuit needed to sample Mandelbrot percolation has constant \depth.  The example of Mandelbrot percolation shows that long range correlations and fractal structures do not entail parallel \depth, all that is needed is the long range communication inherent in the PRAM or circuit model together with arbitrary fan-in gates or, equivalently, concurrent write memory in the PRAM model. The natural decision problem associated with Mandelbrot percolation is in the class \AC\ consisting of problems solvable by circuits of constant depth and arbitrary fan-in.  Note that \AC\ $\subset \NC$.

\subsection{Invasion percolation}
Non-equilibrium cluster growth models based on simple rules and randomness can create fractal patterns without fine tuning a parameter to a critical point.  Examples of systems in this class include invasion percolation~\cite{WiWi} 
and diffusion limited aggregation.  In both of these models, the cluster is initiated as a single particle at the origin and then grows  by the addition of one particle at a time to the
perimeter of the existing cluster. The models differ according to the rules for adding a new particle to the cluster.  Invasion percolation clusters, see Fig.\  \ref{fig:four_fractals}(b) grow on a lattice or graph with weighted nodes.  The weights are identically distributed independent random numbers.  The new particle is added to the site on the perimeter of the existing cluster with the smallest weight.  Invasion percolation models one fluid displacing a second fluid in a porous material, as might occur, for example, as oil is extracted from an oil field by pumping in water.  The black area in the Fig.\  \ref{fig:four_fractals}(b) represents the invading fluid.

For both invasion percolation and DLA,  an essential feature  is that particles are
added {\em one at a time} to the cluster and the location of the next added particle depends on the current shape of the cluster.  It would
seem  that growing a cluster with $N$ particles would then require order $N$ steps even with the help of parallelism.  This conjecture turns out not to be true for invasion percolation.  The task of growing invasion percolation clusters can be transformed to a 
waiting time growth model where sites of the graph or lattice are assigned random waiting 
times from some distribution~\cite{RoHaHi,MaGr96}.  The random waiting times are related to the random weights in the conventional definition of the problem.  A particle is added to a site on the perimeter of the cluster after the site has been on the perimeter for its assigned 
waiting time.  A minimum weight path algorithm~\cite{GiRy} can then be used to find clusters of size $N$ in a
parallel time that is polylog in $N$  showing that the \depth\ of invasion percolation is exponentially
less than  the time required to make clusters according to the sequential defining rules.  

The mapping to a waiting time growth model and parallel solution using minimum weight paths can be applied to other cluster growth models with self-organized critical behavior~\cite{MaGr96}, such as the Eden model~\cite{Eden, RoHaHi} and the restricted solid-on-solid model~\cite{KiKo,TaKeWo}.  For all of these examples, an apparently sequential growth process can be replaced by a parallel growth process that yields the same ensemble of cluster configurations in polylog parallel time.  

The \depth\ of invasion percolation and related models is polylogarthmic in the size of the cluster.  These models have more \depth\ than Mandelbrot percolation but less than might have been expected.  The defining sequential dynamics is replaced by a much more efficient parallel dynamics, which nonetheless produces exactly the same ensemble of configurations.  

\subsection{Diffusion limited aggregation}
\label{sec:dla}

Unlike invasion percolation and its cousins, there is no known way to simulate diffusion limited aggregation efficiently in parallel.  Both the random walk based rule described in Sec.\ \ref{sec:intuitive} and an equivalent rule based on fluid flow in porous
media are associated with \PP-complete problems~\cite{Mac93a, MaGr96}.  The \PP-completeness proof proceeds by a
reduction from the circuit value problem and requires the design of ``gadgets'' that implement
logic gates using the natural dynamical processes of DLA.  Essentially, the proof shows that the
evaluation of an arbitrary circuit can be programmed into the growth of a DLA cluster.  In order to convert the sampling problem to a decision problem, the random numbers that are needed to grow the cluster become the input to the problem.  The \PP-completeness proof shows that, given an arbitrary circuit with specified inputs, we can easily compute the inputs to the DLA problem and then let DLA dynamics compute the output of the circuit. 

The \PP-completeness result suggests that it is
unlikely that there is a way to generate DLA clusters in polylog time.  Nonetheless, some
acceleration is possible using parallelism. DLA generates a tree structure with the particle at the origin serving as the root of the tree.  When a new particle sticks to an existing particle, it becomes the daughter of the existing particle in the tree structure.   It turns out that the main branches of DLA clusters are sufficiently straight that the structural depth of the tree (maximum distance to the origin) scales as the radius of the cluster.   The parallel algorithm of Ref.\ \cite{TiMa04} works by tentatively adding all particles to the aggregate at once and then removing those particles that are obviously in the wrong place because they arrive at their sticking points along a path that cuts across earlier arriving particles.  The algorithm adds nearly one new level to the tree in each parallel step.  Thus, the running time $T$ of the algorithm is proportional to the radius or, in terms of the number of particles in the aggregate, $T \sim N^{1/d_f}$ where $d_f$ is the fractal dimension of the cluster.  For two-dimensional DLA, $1/d_f \approx 0.58$ so the parallel algorithm is more efficient than the sequential algorithm, which cannot do better than $\ord{(N)}$. 

The fact that DLA apparently has greater \depth\ than invasion percolation correlates well with various intuitive
notions of complexity.  In appearance, DLA clusters are more interesting and ``organic'' than  invasion percolation clusters.  While the latter are simple fractals, DLA is described by a
hierarchy of ``multi-fractal'' dimensions.  Finally, invasion percolation is in the same universality class as ordinary percolation, for which there is a
relatively well-developed, controlled theory.  Theoretical analyses of DLA
involve uncontrolled approximations.  

There are several variants of diffusion limited aggregation that deserve mention and differ in their complexity from the standard version.  Internal diffusion limited aggregation (IDLA)~\cite{LaBrGr} is like ordinary DLA except that the walking particles begin at the origin (rather than far from the origin) and stick to the perimeter of the aggregate when they first step off the cluster.  The patterns created by IDLA  are nearly circular.  Instability in the DLA growth process leads to dendritic clusters whereas IDLA growth is very stable.  A perturbation away from the ideal circular shape is quickly damped out because it is more likely for the walker to reach a point closer to the origin.  Intuitively the circular clusters typical of IDLA are less complex than dendritic structures generated by DLA. This intuition is born out by the existence of a parallel algorithm, given in Ref.\ \cite{MoMa00}, that generates IDLA clusters very quickly in parallel.  The algorithm takes advantage of the knowledge that clusters are typically almost circular.  The initial state generated by the algorithm is a circular cluster set up by placing particle $n$ along its walk path at a distance $n^{1/d}$ from the origin, where $d$ is the dimensionality of the system.  This initial state is not an allowed cluster because some particles sit on top of one another and there are also holes in the interior of the cluster.  In the successive steps of the algorithm, particles are moved along their random walk trajectories until all of the holes and multiple occupancies are eliminated.  The average running time of the algorithm was studied by simulating the performance of the parallel algorithm on a single processor workstation.  The data shows that the running time is either polylog or a small power of the number of particles.  For example, it takes, about 10 parallel steps to generate a cluster of 10,000 particles and 13 steps for a cluster of 40,000 particles.   

Although the \PP-completeness proof for ordinary DLA fails for IDLA, it was shown in Ref.\ \cite{MoMa00} that a natural decision problem associated with IDLA is complete for  \CC, the class of problems solvable by polynomial size circuits composed of comparator gates~\cite{GrHoRu}. It is believed that \CC\ and \NC\ are incomparable so that \CC-complete problems are not likely to be solvable by polylog depth circuit families.  On the other hand, it is possible that the sampling algorithm for IDLA runs in polylog time.  These two possibilities are compatible because the \CC-completeness result refers to the worst case and the quoted results for the sampling algorithm refer to the average case. The sampling algorithm takes advantages of the simplifying fact that the typical IDLA cluster is nearly circular.  However it is possible to find atypical instances of IDLA for which the sampling algorithm runs much more slowly than the average. 

A second variant of diffusion limited growth is reversible aggregation (RA)~\cite{DsMa99,DsHoMa2003}.  In this model, particles diffuse and stick to the cluster in the usual way but when they stick, ``heat" particles are liberated into the environment and diffuse.  If heat particles contact ordinary particles on the perimeter of the aggregate, the heat particle is absorbed and the ordinary particle evaporates from the aggregate.    The detailed implementation of RA is, as the name implies, reversible in the sense that the previous state of the system can be obtained from the current state.  In Ref.\ \cite{DsMa99} RA is simulated on a reversible cellular automaton.  Initially there are no heat particles and the aggregate grows much like ordinary DLA.  As the density of heat particles increases, so does the evaporation rate.  The dendritic structure of DLA is gradually replaced by a stringier structure, a branched polymer.  

\subsection{Beyond \PP}
\label{sec:beyondp}
 
Reversible aggregation has greater computational power than ordinary DLA.  The \PP-completeness result for ordinary DLA shows that logic gates can be built using DLA dynamics.  Because sticking events are irreversible, each particle can participate in only one logic operation and the growth of an $N$ particle cluster carries out fewer than $N$ logical operations. Reversible aggregation, on the other hand, allows one to build reusable gates so that each particle can participate in arbitrarily many logical operations~\cite{DsHoMa2003}.  Reversible aggregation is capable of solving \PS-complete problems in contrast to DLA dynamics, which is only known to be capable of solving \PP-complete problems.   

Although RA is associated with a natural \PS-complete problem it does not follow that RA generates exponentially more \depth\ than ordinary DLA.  Reversible aggregation with closed boundary conditions approaches an equilibrium state resembling branched polymers after a reasonably short time.  Although RA has the potential for exponentially long computations, with random initial conditions it almost always settles into an equilibrium state in a much shorter time.  This example illustrates the idea that the potential to solve hard problems is a necessary but not a sufficient condition for great \depth.

Reversible aggregation is one of many examples of cellular automata that have been shown to be ``computationally universal"   in the sense that they are capable of simulating a universal Turing machine.   Another well-known irreversible CA is Conway's Game of Life, which was shown to be computationally universal in~\cite{BeCoGu}.  Wolfram~\cite{Wolf94} has investigated and classified the computational power of cellular automata. 

\NP-complete problems have also been discussed in the physics literature.  One example is the Ising spin glass--a version of the Ising model where each coupling  between
neighboring spins is randomly set to either favor alignment or favor anti-alignment.  More formally, in Eq.\ \ref{eq:isingenergy}, each $J_{ij}$  is independently and randomly chosen to be $+1$ or $-1$ with equal probability.  The
ground state of a spin glass is not simply all spins up or all spins down but depends in a complicated way on the assignment of the $J_{ij}$'s.  Finding the ground state energy of a spin glass in more than two dimensions is \NP-complete~\cite{Bara}. The  \NP-completeness result suggests that
finding ground states requires exponential computational work, there is apparently
no  way to improve dramatically upon an exhaustive search through all $2^N$ states.  Although finding them requires exponential computational work, it is not clear that spin glass ground states are themselves complex in a physical or intuitive sense.  

\section{Discussion}
\label{sec:disc}

In this essay I have introduced a notion of \depth\ for physical systems.  \Depth, defined in the context of statistical physics and parallel computational complexity theory,  characterizes the length of the shortest history, measured in logical steps, needed to generate  typical system states starting from random bits. \Depth\ is defined using the PRAM model, the most powerful, classical model of parallel computation.   Because it discounts hardware and communication costs and is based on the most efficient parallel algorithm, \depth\ is an irreducible measure of history.

We have discussed several examples from statistical physics that show that \depth\ shares features intuitively associated 
with complexity.  For the Ising model,  \depth\ is low for the extremes of high and low temperature where the states are, respectively, highly disordered and highly ordered. It reaches a maximum at the critical point
where correlations in space and time diverge.   The quadratic map provides a rather different illustration of the point that \depth\ reaches a maximum ``at the edge chaos" between periodic and chaotic behavior~\cite{Mac02}.  More generally, for systems where entropy varies with a parameter such as temperature,  \depth, like intuitive complexity,  is expected to reach a maximum somewhere between the highly ordered and highly disordered states. 

We have also examined several non-equilibrium
growth models that illustrate the point that an apparently long history may or may not be an
irreducibly long history.  Diffusion limited aggregation was found to have greater \depth\ than invasion percolation even though both models are defined by rules in which one particle is added to the cluster at a time.  A visual inspection  of patterns generated by these models suggests that DLA is more complex than invasion percolation.  In addition to DLA, invasion percolation and their close relatives,  other non-equilibrium models have been studied from the point of view of \depth.  These include several systems displaying self-organized criticality: sandpiles~\cite{MoNi99}, the Bak-Sneppen model~\cite{MaLi01} and growing networks with preferential attachment~\cite{MaMa05}.  In all of these cases, except sandpile models, efficient parallel algorithms were found that show that \depth\  is much less than might have been expected in systems that spontaneously generate long range correlations in space and time.

The notion of \depth\ can be generalized from system states at a single time to time evolutions or, equivalently, states in space-time~\cite{Mac02,MaLi01}. The \depth\ of a time evolution is the minimum number of parallel steps needed to generate the series of states constituting the time evolution.  The \depth\ of this time series is at least as great as the \depth\ of the deepest state in the series. It may be yet deeper since a simulation of the sequence of states must also reproduce the correlations between these states.  On the other hand, the simulation of the final state in a time series might require producing all of the earlier states, in which case the \depth\ of the final state is the same as the \depth\ of the evolution leading up to it.  

Is \depth\ a physical property?  It has some but not all features of physical properties such as entropy.  \Depth\ is  objective  in the sense that it is the same for all observers and it is an intrinsic property of natural systems whether or not it is measured.  It applies to any system in the purview of statistical physics.  On the other hand, the evaluation of \depth\ is only indirectly related to physical observation and measurement.  Careful model building together with the design and analysis of algorithms are intermediaries between observations and estimates of \depth.  In this regard, \depth\ is unlike, say, entropy, which can be obtained directly by integrating measurements of the specific heat.  It is not surprising that a quantity that captures significant aspects of intuitive complexity requires a long chain of reasoning  and is not directly accessible using simple measuring devices.  

The proposal that \depth\ is a useful proxy for complexity lacks content unless it could be falsified.  There are two main ways this could happen.  On the one hand, \depth\ might be too weak a requirement for complexity. It may be that most natural systems, when simulated with sufficient realism, have substantial \depth.  In Sec.\ \ref{sec:intuitive} we speculated that the Earth, with its biosphere, had considerably more \depth\ than the Sun but this speculation and others like it may not hold. \Depth\ may fail to distinguish the somewhat complex from the very complex.  On the other hand, \depth\ could prove too strong a requirement for complexity.  If there are systems that are clearly very complex but have little \depth\ then the hypothesis that an irreducibly long history is needed for complexity is incorrect.  If it turns out that any feasible simulation can be efficiently parallelized, e.g. if $\PP=\NC$, then it would follow that no physical process that can be simulated in polynomial time would have much \depth.

If  \depth\ or another proxy for complexity gains wide acceptance, it opens the way for asking a number of interesting questions that go beyond the analysis of the simplified models discussed here. We could ask what features of natural systems either encourage or suppress
complexity and why biological evolution so efficiently increases complexity.
On a cosmic scale, \depth\ might provide an alternative to the anthropic principle~\cite{LiRe05}.  The anthropic principle offers a way to understand why certain apparently arbitrary physical
parameters have their realized values by requiring that these parameters are compatible with the
existence of intelligent life.  Since we have little understanding of what conditions are
required for intelligent life to exist, real applications of the anthropic principle look for
simple and ad hoc preconditions for life as we know it such as a reasonably long life-span for the universe and the existence of elements beyond hydrogen.  An alternative to the anthropic principle might be the requirement that the laws of physics and the undetermined fundamental parameters are set so that the universe can develop substantial \depth.

\section{Acknowledgments}
This work was supported in part by NSF grant DMR-0242402.  I am indebted to  Ray Greenlaw, Neil Immerman, Ben Machta, John Mayfield and Cris Moore for many useful discussions and for their comments on the manuscript.

\end{document}